\documentclass[twocolumn, tradiabstract]{aa} 

\def\rmit#1{{\it #1}}              
\def\specchar#1{{\sc #1}}

\def\eg{\rmit{e.g.}}
\def\degree{\hbox{$^\circ$}}
\def\arcsec{\hbox{$^{\prime\prime}$}}
\def\FeI{\mbox{Fe\,\specchar{i}}}
\def\FeII{\mbox{Fe\,\specchar{ii}}}

\def\kms{\hbox{km$\;$s$^{-1}$}}
\def\ms{\hbox{m$\;$s$^{-1}$}}

\usepackage{natbib}
\usepackage{graphicx}
\usepackage{color}
\usepackage[normalem]{ulem}

\begin{document}

\title{Evershed flow observed in neutral and singly ionized iron lines}

\author{E. Khomenko\inst{1,2}, M. Collados\inst{1,2}, N. Shchukina\inst{1,3},  A. D\'{\i}az\inst{4} }

\institute{Instituto de Astrof\'{\i}sica de Canarias, 38205 La Laguna, Tenerife, Spain \and 
Departamento de Astrof\'{\i}sica, Universidad de La Laguna, 38205, La Laguna, Tenerife, Spain \and 
Main Astronomical Observatory, NAS, 03680, Kyiv, Ukraine \and
Departament de F\'{\i}sica, Universitat de les Illes Balears, E-07122 Palma de Mallorca, Spain\\ \email{khomenko@iac.es} \\
}

\date{Received XXX, 2015; accepted xxx, 2015}

\abstract{The amplitudes of the Evershed flow are measured using pairs of carefully selected \FeI\ and \FeII\ spectral lines located close in wavelength and registered simultaneously. A sunspot belonging to the NOAA 11582 group was scanned using the spectrograph of the German Vacuum Tower Telescope (Observatorio del Teide, Tenerife). Velocities were extracted from intensity profiles using the $\lambda$-meter technique. The formation heights of the observed spectral lines were calculated using semi-empirical models of a bright and dark penumbral filament taking into account the sunspot location at the limb. Our objective is to compare azimuthally averaged amplitudes of the Evershed flow extracted from neutral and ion lines. We find measurable differences in the radial component of the flow. All five pairs of lines show the same tendency, with a few hundred \ms\ larger amplitude of the flow measured from \FeI\ lines compared to \FeII\ lines. This tendency is preserved at all photospheric heights and radial distances in the penumbra. We discuss the possible origin of this effect.}


\authorrunning{Khomenko et al.}
\titlerunning{Evershed flow from \FeI\ and \FeII\ lines}

\maketitle

\section{Introduction}

The Evershed flow, discovered more than a century ago, was detected as a nearly horizontal outflow in sunspot penumbrae at photospheric heights, with a typical speed of a few \kms\ \citep{Evershed1909, Bray+Loughhead1979}. The Evershed flow speed decreases with height and changes sign somewhere in the upper photosphere, such that an inflow is observed at chromospheric heights \citep{StJohn1913}. Early works detected that the radial velocity component has a maximum at some distance in the penumbra close to the umbra border, and that the flow speed decreases toward the end of the penumbra \citep{Kinman1952}, but continue beyond its outer end.

It is nowadays understood that there is a strong relationship between the properties of the Evershed flow and the penumbral fine structure. The sunspot penumbra, when observed at high spatial resolution, is a highly inhomogeneous medium. The magnetic field vector derived from observations in dark and bright penumbral filaments shows strong spatial variations. The magnetic field in penumbra is structured in regions with stronger and more vertical field, usually called spines \citep{Lites+etal1993, Langhans+etal2005}. In between, there are regions with more horizontal and weaker field (intra-spines).  On average, the inclination of the magnetic field with respect to the vertical gravity direction changes between 40 and 80 degrees from the inner to the outer penumbra. The difference between the inclination of spines and the nearby intra-spines at the same radial distance reaches about 10-30 degrees.  Such a structure of the penumbral magnetic field is often called uncombed penumbra \citep{Solanki+Montavon1993}. Because of the total pressure balance that has to be maintained in horizontal direction, the surface level in intra-spines must be elevated with respect to spines \citep{Borrero_LR}.  

The Evershed flow is not uniformly distributed over the penumbra but is confined in the narrow channels of inter-spines with more horizontal field \citep{Stanchfield+etal1997, Mathew+etal2003}. The relation between the brightness of penumbral filaments and the Evershed flow is not straightforward. There are indications that the more horizontal component of intraspines coincides with dark filaments in the outer penumbra and with bright ones in the inner penumbra \citep{Schlichenmaier+etal2005, BellotRubio+etal2006, Ichimoto+etal2007}.  Up- and downflows are superimposed on the radial outflow and produce a difference in the velocity inclination angle between the bright and dark features, depending on the radial distance in the penumbra \citep{Schmidt+Schlichenmaier2000}. Based on the inversion of Stokes profiles of the infrared \FeI\ lines at 1.56$\mu$m, \citet{BellotRubio+etal2003} established that the magnetic field and the velocity of the Evershed flow were aligned all over the inner and outer penumbra, when considering a two-component inversion. This result can be taken as a confirmation that the magnetic field is frozen into plasma in sunspot penumbra. 

Observations of the Evershed flow were done using photospheric lines, often iron lines, forming rather deep down in the solar atmosphere. At those heights, the collisional coupling of the plasma is strong and the frozen-in condition can be reasonably expected, when considering a fully ionized plasma.  However, the photosphere is only weakly ionized and the ionization fraction becomes even lower in the cool atmospheres of sunspots, see \citet{Khomenko+etal2014b}. Even though the collisions are strong in the photosphere, it can be expected that ions may become magnetized at some height in the photosphere because the magnetic field in sunspots is very strong. This may cause partial decoupling of the neutral and charged species. The decoupling may produce differences in their velocities, since different forces are acting on them.  One the one hand, due to collisions between neutral and charged particles (the latter being tied to the magnetic field), the magnetic field can diffuse through the plasma, a process often called ``ambipolar diffusion'' in the astrophysical context. This may lead to a partial break of the frozen-in condition. On the other hand, there can appear additional forces associated to misbalance of the partial pressure gradients of different species.  Therefore, in principle, the velocities measured from spectral lines of neutral and ionized species in sunspots may slightly differ.  Up to our knowledge, there have been no attempts to compare the properties of the flow from simultaneous measurements of neutral and ion spectral lines. Measurements done separately, i. e. in different sets of observations by different authors, do not reveal qualitative differences in the behavior between these species. 

Direct observational confirmation of the uncoupled behavior of neutral and ionized species in the solar atmosphere is still missing. Order of magnitude calculations show that the scales at which ion-neutral effects become important are rather small. At maximum, spatial scales reach few kilometers, and temporal scales reach fraction of seconds, depending on the height in the atmosphere and on the magnetic field strength \citep{Khomenko+etal2014b}. Direct observation of such small scales is not possible with current observational facilities. Nevertheless, if the differences in the behavior of ions and neutrals are sufficiently large in a small area inside the resolution element, it may be possible that their traces remain in the integrated signal 
even at the typical resolution of observations. This gives a chance of their detection in a specially dedicated observational campaign.  Below we describe the results from such campaign targeting the Evershed flow in sunspots.

\begin{center}
\begin{table*}
\begin{center}
\caption{Parameters of the observed spectral lines. } 
\begin{tabular}{cccccccc}
\hline Elm Ion & $\lambda$ (\AA\ )  &  EPL (eV) & $\log gf$ & $g_{\rm eff}$ & $H_C$(km) & $H_D$(km) & I \\ \hline
\FeII\ & 4576.34  & 2.84  & $-3.05$ & 1.18  &  448  & 292  &  0.69\\
\FeI\  & 4574.22  & 3.21  & $-2.44$ & 1.90  &  280  & 269  &  0.48\\
\hline
\FeII\ &  4656.98 &  2.89 & $-3.80$ &  1.67  & 163 & 171 &  0.41\\
\FeI\  &  4657.58 &  2.85 & $-2.95$ &  1.40  & 277 & 273 &  0.41\\
\hline
\FeI\  & 5196.06  & 4.26 & $-0.83$ &  1.09  & 336  & 350 &  0.69\\
\FeII\ & 5197.58  & 3.23 & $-2.38$ &  0.67  & 538  & 328 &  0.72\\
\FeI\  & 5197.94  & 4.30 & $-1.52$ &  0.36  & 216  & 207 & 0.39\\
\FeI\  & 5198.71  & 2.22 & $-2.09$ &  1.50  & 398  & 494 & 0.79\\
\hline
\FeII\ & 5234.62 & 3.22 & $-2.31$ &  0.87 &  553   & 331 & 0.72\\
\FeI\  & 5236.20 & 4.19 & $-1.75$ &  0.39 &  235   & 244  & 0.38\\
\hline
\FeII\ & 6516.08 &  2.89 & $-3.43$ &   1.07 &  279 & 250 & 0.46\\
\FeI\  & 6518.37 &  2.83 & $-2.64$ &   1.15  & 327 & 351 & 0.51\\
\hline
\end{tabular}
\end{center}
\end{table*}
\end{center}

\section{Observations and data analysis}

The observations were done at the German Vacuum Tower Telescope (VTT) at the Observatorio del Teide (Iza\~na, Tenerife) during the morning of 7th of October, 2012. As a target we used an isolated regular-shaped sunspot belonging to the NOAA 11582 group located at coordinates (914\arcsec,  -216\arcsec) off solar disc center.  By means of the spectrograph of the VTT we scanned the sunspot using several carefully selected spectral intervals.  Each interval was selected from the the list of unblended spectral lines with a clean continuum provided in \citet{Gurtovenko+Kostik1989} under the criterion of containing at least one \FeI\ and one \FeII\ spectral line close in wavelength to fit on the CCD. We selected pairs of lines with as similar formation heights as possible. The lines are listed in Table 1. The columns of the Table from left to right give the ionization stage, wavelength  $\lambda$, lower level excitation potential EPL,  logarithm of the oscillator strength times the multiplicity of the lower level log$gf$, effective Land\'e factor $g_{\rm eff}$, line minimum formation height $H_C$ and $H_D$ and central line depth $I$. The effective Land\'{e} factors $g_{\rm eff}$ are evaluated using the L-S coupling scheme. All of the spectral lines in the list have a moderate sensitivity to the magnetic field. Atomic data are taken from \citet{Gurtovenko+Kostik1989}. The line core formation heights $H_C$ and $H_D$ are calculated in the penumbral models C and D from \citet{Socas-Navarro2007}, see Section \ref{sect:formheights} below. 

The observed dataset also contains the lines of other elements, as Ti and Ba, forming higher up in the atmosphere. However, in that case, only one spectral line was scanned at a time. Since we perform a comparative analysis of ion and neutral velocities and we do not expect the differences to be large, we excluded those lines from the current analysis. Uncertainties derived from the errors in the wavelength calibration, or simply temporal changes in the properties of the Evershed flow, may influence the results of the comparison.

\begin{figure}[!]
\center
\includegraphics[width=8.0cm]{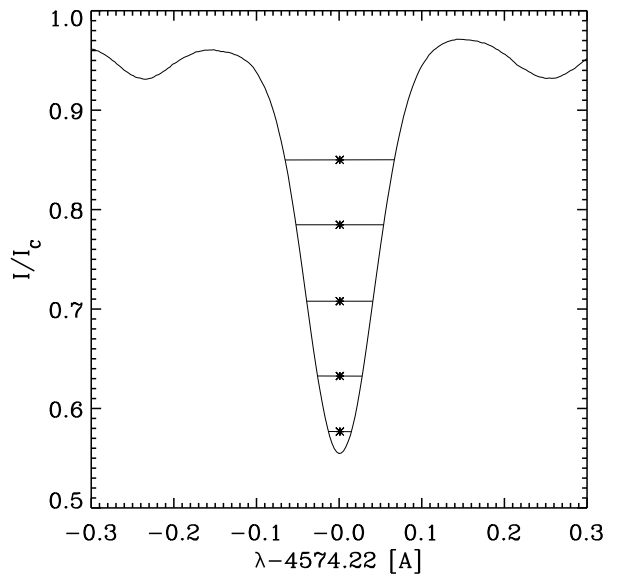}
\caption{Average over the observed field of view profile of the \FeI\  4575.22 \AA\ line. The horizontal lines show the positions of the five chords  of fixed width used for the determination of the $\lambda$-meter velocities. The velocity corresponding to each of the chords is calculated as the mid-point between the blue and the red wings and is represented by the symbol. }
\label{fig:example_lm}
\end{figure}

The different spectral regions were scanned successively starting at 9:26 UT. The scan step was 0.18\arcsec\ and 250 positions were taken. It took about 5 min to complete each scan. The seeing conditions were exceptional during the morning, with the Fried's $r_0$ parameter varying around 15-20 cm and the VTT adaptive optics system \citep{KAOS} locking all the time on the observed sunspot. In fact, some of the observed \FeI\ and \FeII\ lines showed emission on the limb, that is rarely detected in the measurements done from the ground due to the influence of the Earth atmosphere \citep[see e.g.,][]{Lites+etal2010}. This marks the exceptional quality of the observational dataset.

The data were reduced using a standard procedure. The reduction includes the correction for the curvature of spectral lines on the camera.  The velocities were extracted by means of the $\lambda$-meter technique \citep{Stebbins+Goode1987, Shchukina+etal2009, Kostik+etal2009}. According to this technique we selected five chords of fixed width for each spectral line, based on its spatially averaged profile over the whole field of view. We then calculated the blue-wing and red-wing displacements corresponding to each chord and spatial position. The mid-points between blue- and red-wing displacements were taken as velocities corresponding to a given height and spatial location. Figure \ref{fig:example_lm} shows an example of the average over the whole field of view profile of the \FeI\ 4575.22 \AA\ line, together with five $\lambda$-meter chords and symbols indicating the mid-point displacements (very close to zero for the average profile).  The spectral lines used in this study have rather low Land\'e factors and are not visibly Zeeman-split in the penumbra. 
Therefore, the adopted method, to first order, does not suffer from errors derived from fitting a chord into profiles having several distinct Zeeman components. We assigned a formation height to each chord using the procedure described in the section below. 

As discussed in the Introduction, the Evershed flow is not distributed uniformly over the penumbra, but rather propagates in the narrow channels corresponding to the more horizontal magnetic field component of intra-spines. The line profiles resulting from not completely resolved penumbral filaments, as in the case of our data, are better represented by a two-component model, where each component has different magnetic field strength, inclination and velocity. Unfortunately, in our dataset, we have no polarimetric information and there is not enough information in just the intensity profiles of the observed spectral lines to extract the information about two magnetic components of the penumbra. Therefore, instead of performing inversions, we rather applied the $\lambda$-meter simplified procedure to extract an average velocity field over the different components in the penumbra at each spatial location. We discuss the drawbacks of this procedure in the last Section.

No absolute wavelength calibration was possible due to the absence of telluric spectral lines in the observed spectral intervals. We performed the wavelength calibration by fitting the observed average quiet Sun spectrum (outside of the umbra and penumbra) to the FTS atlas. Since in all the cases the pairs of \FeI\ and \FeII\ lines fit on the same CCD, the uncertainty introduced by such wavelength calibration is the same for both spectral lines. 

\begin{figure*}[!]
\center
\includegraphics[width=18.0cm]{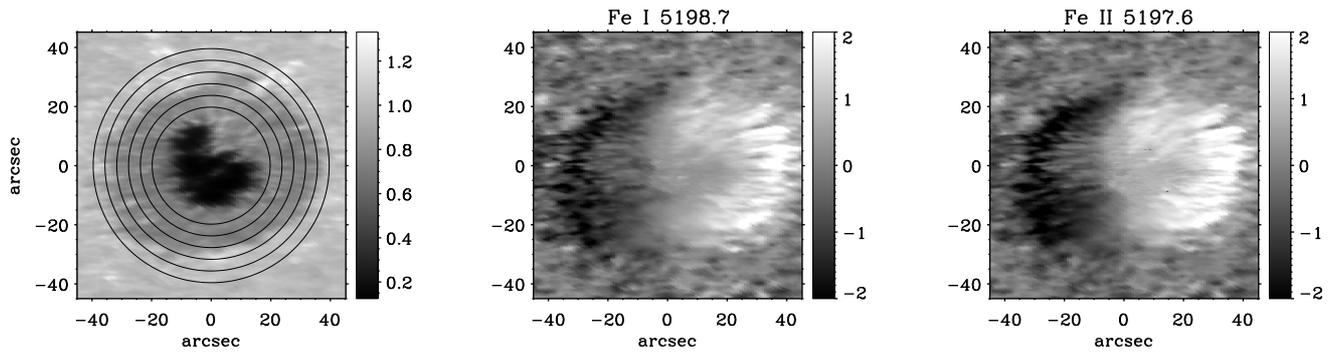}
\caption{Continuum image and velocities of the \FeI\ 5198.71 \AA\ and \FeII\ 5197.58 \AA\ lines. The velocities are obtained after the $\lambda$-meter technique at the position at the wings of each line nearest to the continuum. Notice the similarity in the velocity field. The circles on the left panel mark the rings in the penumbra where we performed the fit to the velocities.}
\label{fig:contin}
\end{figure*}

The 2D images of $\lambda$-meter velocities and continuum intensity were rotated and corrected for the limb perspective. Examples of the continuum intensity and velocities of the \FeI\ 5198.71 \AA\ and \FeII\ 5197.58 \AA\ lines are given in Figure \ref{fig:contin}. There is a large similarity in the velocity field derived from both lines. A visual comparison of both velocity maps reveal differences in the amplitude of the flow only at small-scale details. However, these images can not be compared point to point, since the velocities correspond to slightly different heights. Rather than performing a comparison of the spatially resolved velocities, we decided to perform a statistical analysis and to consider their azimuthal variation. For that we followed the procedure introduced by \citet{Schlichenmaier+Schmidt2000} \citep[see also][]{BellotRubio+etal2003} and inferred azimuthally averaged velocities.

\begin{figure*}[!]
\center
\includegraphics[width=15.0cm]{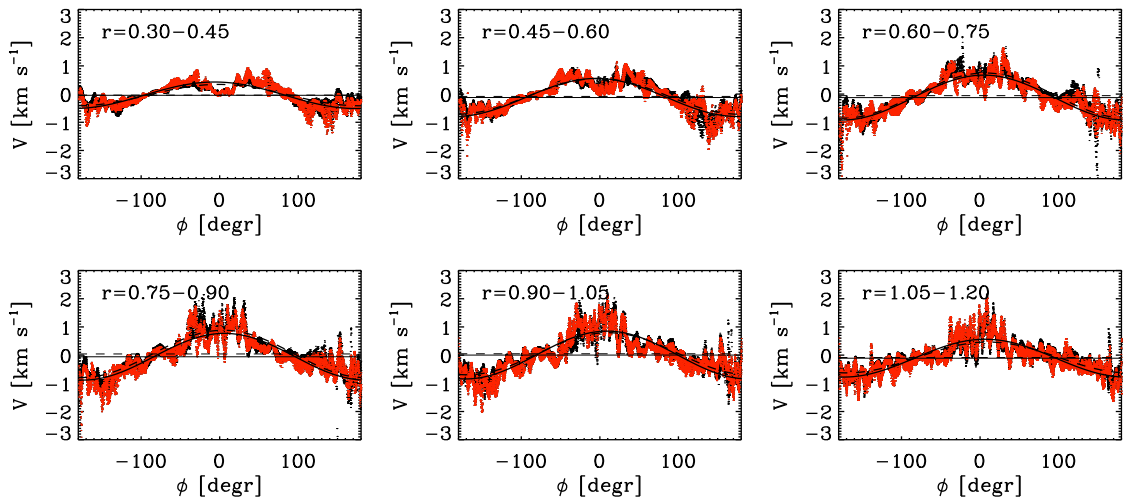}
\caption{Example of the velocity variation in different rings of the penumbra, together with the results of the fit. Red color:   \FeI\ 5198.71 \AA\  line (dashed line for the fit), black color: \FeII\ 5197.58 \AA\ line (solid line for the fit). Horizontal dashed and solid lines give the average velocity values. Velocities are $\lambda$-meter velocities corresponding to the position close to the line core for both lines, $\lambda_1$. Note that the velocities from \FeI\ and \FeII\ lines correspond to different heights.}
\label{fig:vlambda}
\end{figure*}

\begin{figure*}[!]
\center
\includegraphics[width=8.0cm]{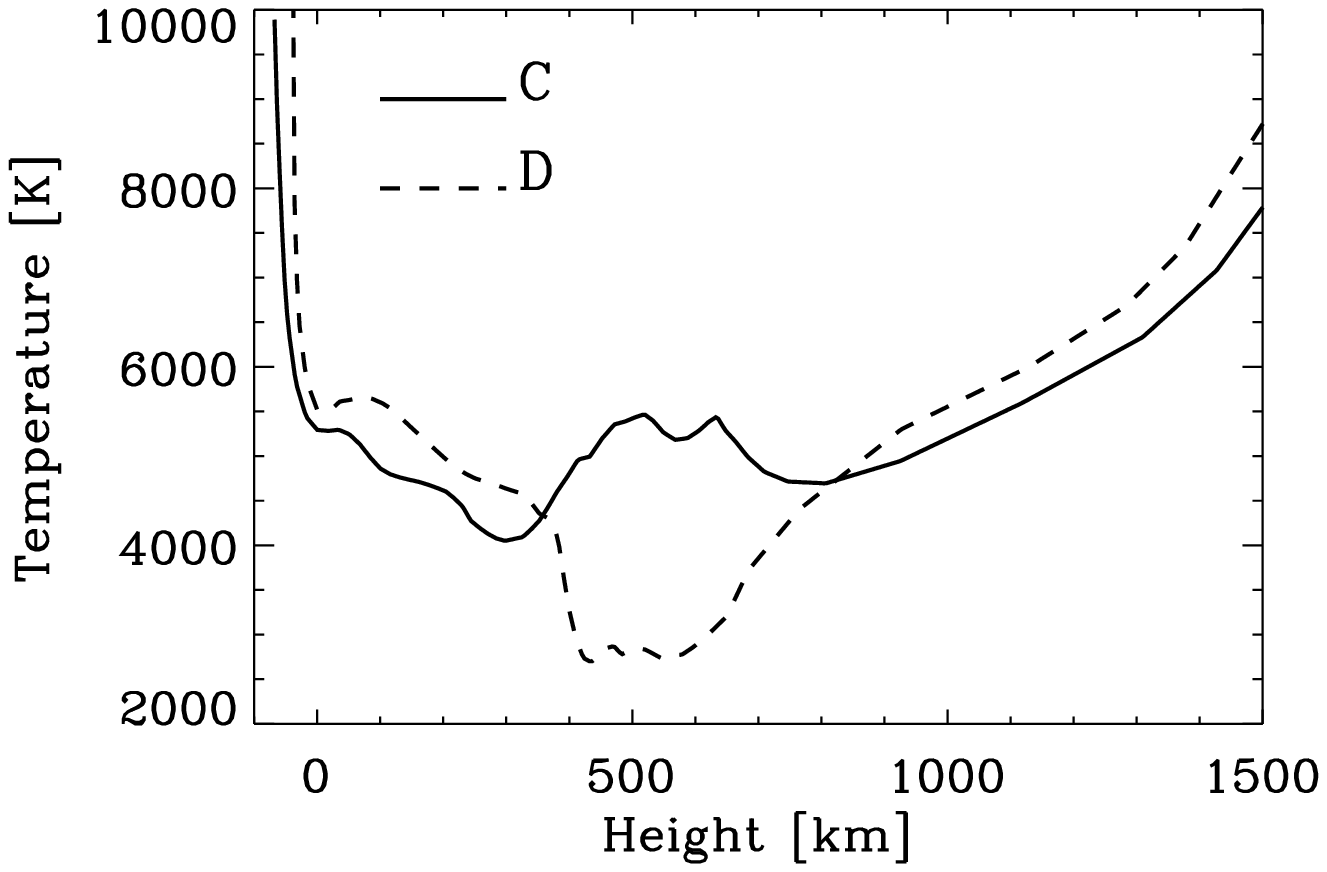}
\includegraphics[width=8.0cm]{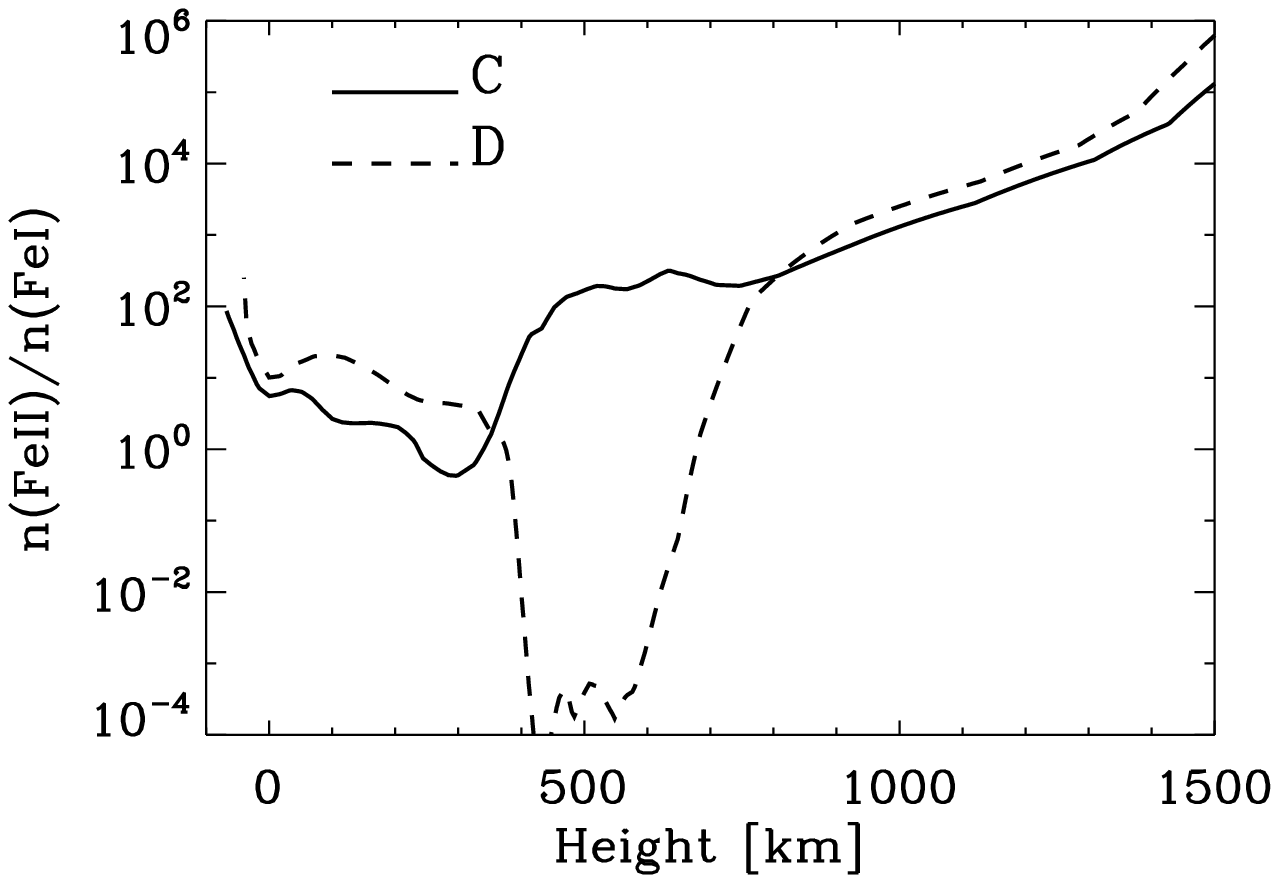}
\caption{Temperatures (left) and relative number densities of the ionized and neutral iron (right) as a function of height in the models C and D used for the calculation of the formation heights. }
\label{fig:n-fe}
\end{figure*}

We selected the LOS velocities corresponding to different radial distances in the penumbra, starting from $0.3R$ (where $R$ is the radius of the penumbra defined ``by eye'' from the continuum image) and finishing at $~1.2R$, see the left panel of Fig.~\ref{fig:contin}. Velocities in the corresponding intervals of radial distances, $0.3-0.45R$, $0.45-0.6R$, $0.6-0.75R$, $0.75-0.9R$, $0.9-1.05R$, $1.05-1.2R$ were fit by the following function: 
\begin{eqnarray}
v_{\rm LOS}(r, \phi) &=& v_x\sin\theta\cos\phi + v_y\sin\theta\sin\phi + v_z\cos\theta \\ \nonumber
&=& v_r\sin\theta\cos(\phi+\alpha)+ v_z\cos\theta
\end{eqnarray}
where $r$ refers to the radial distance interval, $\theta$ is the heliocentric angle, $\phi$ is the azimuthal angle around the spot center, and $v_x$, $v_y$ and $v_z$ are the horizontal and vertical components of the velocity to be determined by the fit. Alternatively the fit can be expressed in terms of radial velocity $v_r$ and the phase $\alpha$. The velocities given by the fit are in the local reference system related to the gravity direction and the axis coinciding with the sunspot umbra center. The fit was done separately for each $\lambda$-meter velocity set corresponding to five chords ($\lambda_1-\lambda_5$, where index 1 corresponds to line core and 5 to line wing) of each spectral line. Figure ~\ref{fig:vlambda} illustrates the $\lambda$-meter velocities corresponding to a position close to the line core of the \FeI\  5198.71 \AA\ and \FeII\ 5197.58 \AA\ lines, as a function of azimuth $\phi$  for different radial positions, together with the fit.

\subsection{Calculation of velocity errors}

We assumed that the uncertainties of the velocity measurements come essentially from the determination of $\lambda$-meter velocities from the line profiles. Such errors are limited by the wavelength resolution of observations. We took an upper estimate of such errors of a half of a pixel size in wavelength. The pixel size varies between 160--180 \ms, depending on the spectral region. Given the error estimate of line of sight velocity measurements, we used the error propagation rule to calculate the errors of vertical and radial velocities and that of the inclination angle of the flow. 

\begin{figure*}[!]
\center
\includegraphics[width=17.0cm]{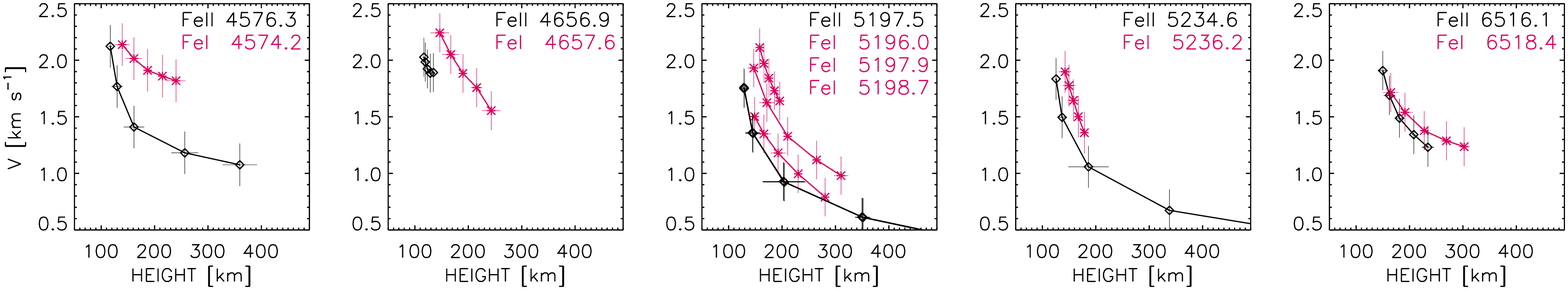}
\includegraphics[width=17.0cm]{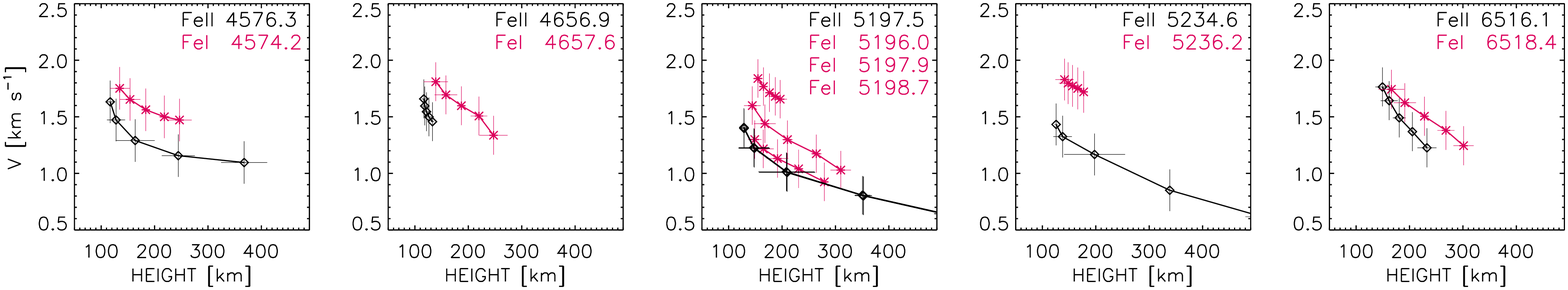}
\caption{Height dependence of the radial component of the Evershed flow obtained after the fit to $\lambda$-meter velocities for different pairs of lines at difference radial distances in the penumbra. The model for heights is model C. Upper row corresponds to the radial distances $0.45-0.6R$, while bottom row corresponds to $0.75-0.9R$.}
\label{fig:vr_lambda_c}
\end{figure*}

\begin{figure*}[!]
\center
\includegraphics[width=17.0cm]{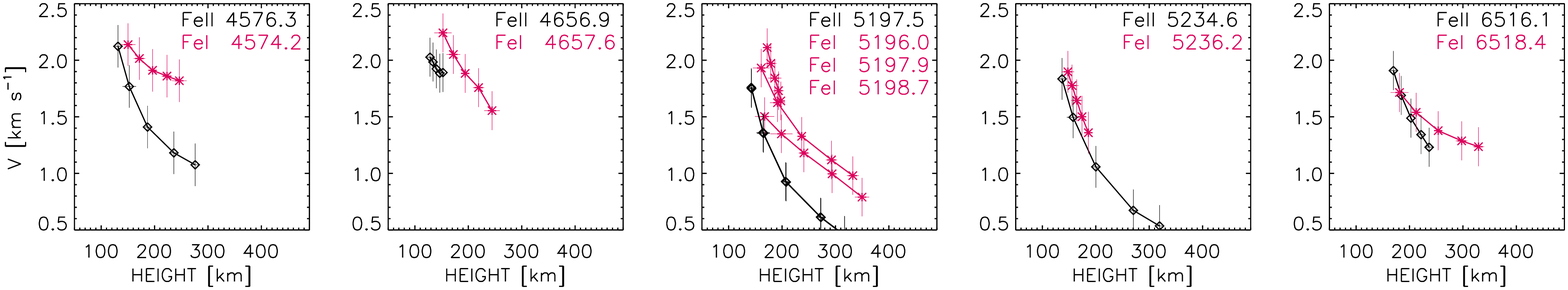}
\includegraphics[width=17.0cm]{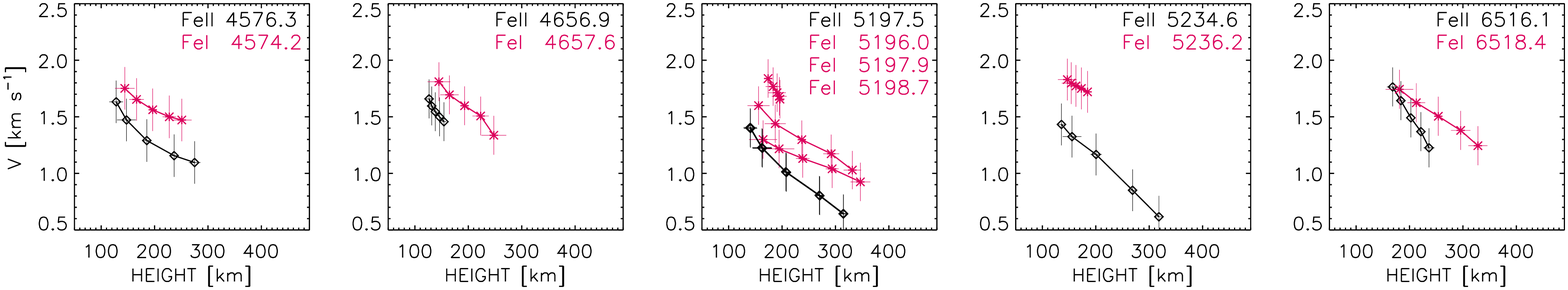}
\caption{Same as Fig. \ref{fig:vr_lambda_c} but for the heights from model D. }
\label{fig:vr_lambda_d}
\end{figure*}

\section{Formation heights of spectral lines}
\label{sect:formheights}

The $\lambda$-meter technique allows to obtain the velocity at several positions along the line profile that correspond to different heights in the solar atmosphere. In order to compare \FeI\ and \FeII\ velocities, we need these heights to be reliably calculated. The formation height of spectral lines  can strongly vary at the locations of bright and dark penumbra filaments. Also, the specific location of the observed spot at the limb should be taken into account. We calculated the formation heights of the observed spectral lines at the observed heliocentric angle of $\theta=77$\degree\ using two semi-empirical models of sunspot penumbra, corresponding to a dark filament (model C) and a bright filament (model D) as provided in \citet{Socas-Navarro2007}. We took into account the magnetic field from the models producing Zeeman splitting or broadening  of the spectral lines, but we did not take into account the velocity field. We have used a radiative transfer code based on the DELOPAR method proposed by \citet{TrujilloBueno2003}, which allows us to compute the emergent Stokes profiles taking into account the Zeeman effect.
The line intensity profiles and formation heights were calculated using the oscillator strengths and effective Land\'{e} factors, $g_{\rm eff}$, given in Table 1. 

Figure \ref{fig:n-fe} shows the distribution of temperatures and the relative number densities of \FeII\ to \FeI\ atoms in the models C and D. The temperatures in these models are strongly different in the middle photosphere. It has been noted by \citet{Shchukina+TrujilloBueno2001} that the ionized iron is the dominant state over the quiet photosphere both in hotter granules and cooler inter granular lanes. In the case of model D, its cooler temperature around 500 km causes the reversal of this trend and produces a larger number of \FeI\ atoms compared to \FeII\ ions. Such strong variations significantly affect the formation heights of spectral lines, as can be observed from Table 1. 

When synthesizing the iron lines, we had in mind that non-LTE effects mainly affects their line source $S_L$ function, but no the line opacities \citep[see, \eg,][and references therein]{Shchukina+TrujilloBueno1997, Shchukina+TrujilloBueno2001}. To simplify the non-LTE modeling of the \FeI\ and \FeII\ lines we considered a schematic atomic model consisting of two levels, between which only radiative transitions can occur, while collisional transitions were neglected. This model provides a more realistic description of the line core intensity compared to the case of pure LTE. In LTE some of the considered spectral lines unrealistically go into emission at the observed limb position. In our approximation of a two-level atom, the line source function $S_L$  was taken equal to the mean intensity $\langle J \rangle$ averaged over the line profile \citep[see][]{Mihalas}. For simplicity, we neglected the non-LTE opacity shift.

The formation heights of the \FeI\ and \FeI\ lines have been estimated using the concept of ``Eddington-Barbier height of line formation''; i.e., we calculated the height where the line optical depth at a given line wavelength point ${\Delta \lambda}$ is equal to unity.

The formation heights of the $\lambda_i$ chords of each spectral line were determined separately for every observed pixel. For that, we determined the wavelengths where the chord crosses the left and the right wing of the observed line profiles. Those wavelength have, in general different formation heights because of the existence of the velocity field in observations. We took the average formation height between the left and right wavelengths as the formation height of each chord. The spatial variation of the formation height from point to point in the penumbra was not strong, of a few km. Since we perform a fit at a given radial position for determining velocities, we also averaged heights in each individual radial ring. We took the standard deviation between the mean and the individual values inside each ring as a measure of the uncertainty of the height determination. The errors defined this way are shown in Figures \ref{fig:vr_lambda_c}, \ref{fig:vr_lambda_d} , \ref{fig:vz_lambda_d} and \ref{fig:theta_lambda_d} as horizontal bars attached to each point. The variation of the formation height with $R$ were not significant. We repeated the above calculation of the formation heights separately for the models $C$ and $D$.

\section{Velocities of \FeI\ and \FeII\ lines}

Figures \ref{fig:vr_lambda_c} and \ref{fig:vr_lambda_d}  show the radial component of the Evershed flow velocity $v_r$ as a function of height in the models C and D for all five spectral intervals separately for the \FeI\ lines (red) and \FeII\ lines (black). 
We only  analyze the results for the radial distances from 0.45 $R$ to 1.05 $R$ of visible penumbral. We discarded the results from the inner penumbra, because the determination of velocities is affected by the Zeeman splitting of spectral lines. We also discarded the results at locations outside of the visible penumbra since the flow speed is small and is affected by the errors of velocity determination. Due to the location of the observed sunspot at the limb, the radial component of the flow is the parameter that is more reliably determined from the fit.  As follows from Fig. \ref{fig:vr_lambda_c} and \ref{fig:vr_lambda_d}, all considered spectral lines show similar dependence of the flow speed with height. The speed decreases from about 2 \kms\ at 100 km to below 1 \kms\ at 400 km. The radial dependence of the flow is also as expected: the values are higher at the central penumbra  (first rows), decreasing towards its end.  

All spectral intervals show the same tendency in the magnitude of the ion and neutral radial velocity component. Its values are few hundred \ms\ larger for the neutral  lines than for the ionized lines. For the heights calculated in the model D, some of the spectral lines (4575 \AA,  at 5197 \AA, and 6517 \AA) show a tendency that the difference between ion and neutral velocities increases with height at the central penumbra (first row in Fig. \ref{fig:vr_lambda_d}). In the other two cases (4657 \AA\ and 5324 \AA) one of the lines in the pair has a very deep formation height to appreciate the difference. In the case of the heights from model C (Fig. \ref{fig:vr_lambda_c}) this tendency is less evident. Even though the difference is slightly above the error bars, the result that neutral velocities are larger than ion velocities is independent from the model used for the calculation of heights. In the case of the 5197 \AA\ spectral interval, all neutral lines measured simultaneously have larger velocities than the ionized line. Nevertheless, there is a difference between the neutral velocities themselves, of the order of the difference between the ion and neutral ones. The difference between the velocities from neutral lines at  5197 \AA\ becomes smaller when using heights from model D. While the results of our analysis are subject to uncertainties due to the height of formation of spectral lines, it is intriguing that all spectral intervals independently show a similar behavior with a slightly larger velocity for neutral iron at the same photospheric heights. The result is also relatively independent of the model used for the calculation of formation heights, which gives it additional validity.

The vertical component of the Evershed flow is determined significantly poorer from the fit. This component is given by the average of the radial variation of the observed LOS velocity at the given radius and is affected by the non-perfect velocity calibration. During the process of the data reduction we removed the curvature of the spectral lines on the camera, produced by the the spectrograph, that may result in false velocity variations over the observed field of view. Apart from that, there is a gradient due to the solar rotation at the time of observations. The magnitude of the velocity change due to solar rotation is about 100 \ms. This gradient was also removed. However, due to the small field of view and the presence of the sunspot,  some residual errors may remain after its removal. Finally, the $\lambda$-meter method defines velocities as the central point of the bisector. This point may have a net shift since the average bisector is not a straight line. When the spectral lines form in solar granulation, the bisector shows a typical C shape caused by the correlations of the velocity and intensity fields  \citep{Kostik1977, Dravins1982}. In the penumbra the shape of the average bisector is also affected by the correlations between the velocity, intensity and magnetic fields \citep[see e.g.,][]{Wiehr+etal1984, Ichimoto1988}, with the asymmetry in the same direction as the displacement. Apart from the asymmetry, an uncertainty could also come from different convective blueshift of neutral and ionized lines similar to the quiet Sun, if one assumes that convection is present in the penumbra. Additionally, the shape of the bisector of each line can be affected by blends in the wings (in some of the cases). The shape changes from line to line. In order to correct for this effect, we subtracted the position of the average bisector over the whole penumbra from the $\lambda$-meter velocities.

\begin{figure*}[!]
\center
\includegraphics[width=17.0cm]{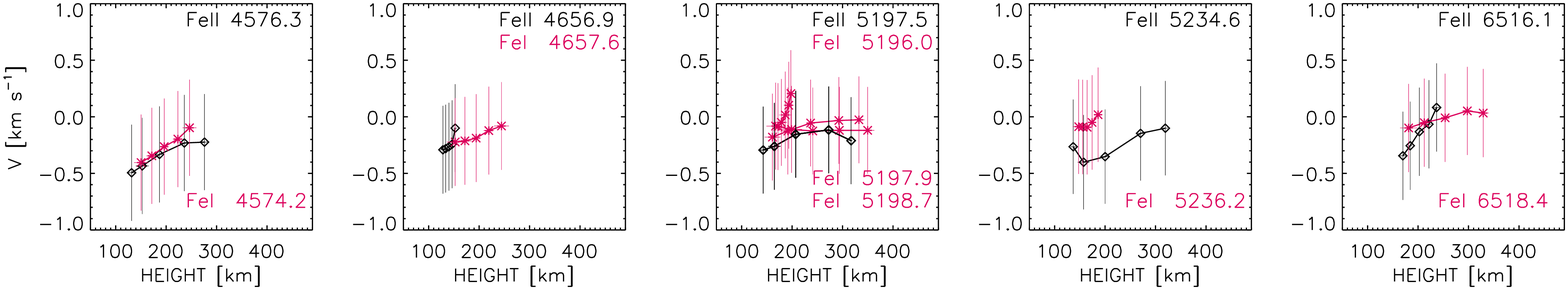}
\includegraphics[width=17.0cm]{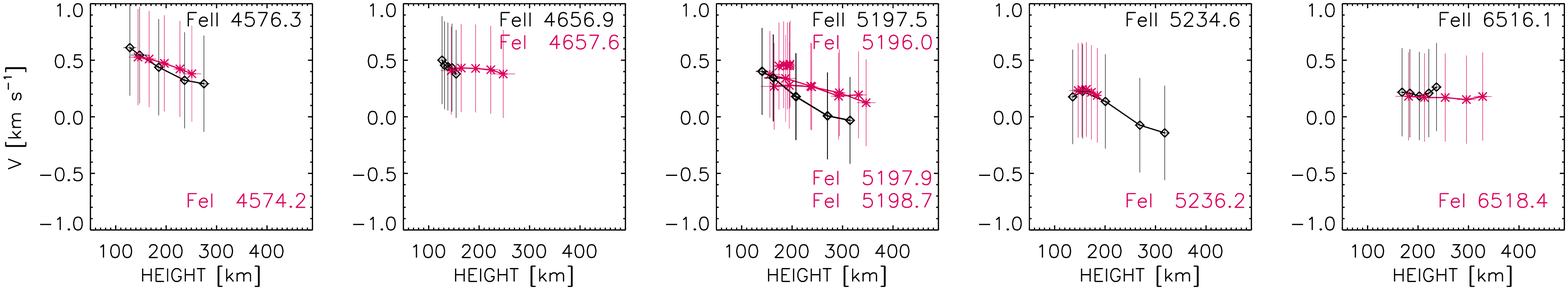}
\caption{Same as Fig. \ref{fig:vr_lambda_d} but for the vertical component of the Evershed flow. The heights are from model D. }
\label{fig:vz_lambda_d}
\end{figure*}

The resulting vertical components of the velocities are given in Figure \ref{fig:vz_lambda_d} for the heights from model D. Even after all corrections, the error bars are too large to reliably determine if there is difference between the ion and neutral velocities. Apparently, the observations do not show any tendency, as was the case for radial velocities. The difference in all cases remains within the error bar limit. We only show the results for one model of heights, since the results for the other one are qualitatively similar.  

\begin{figure*}[!]
\center
\includegraphics[width=17.0cm]{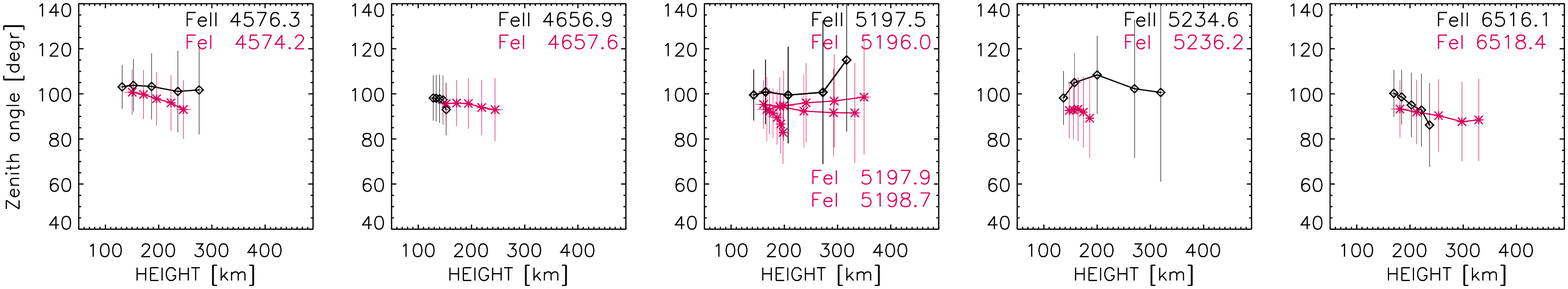}
\includegraphics[width=17.0cm]{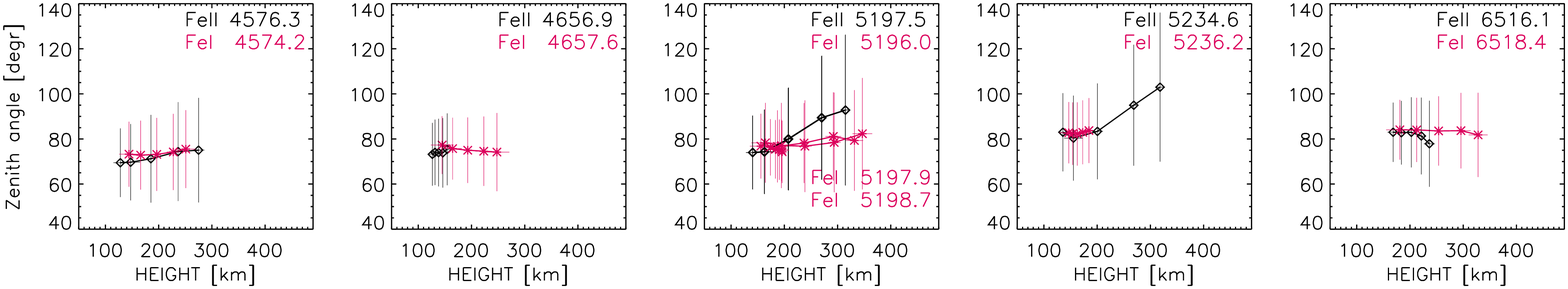}
\caption{Same as Fig. \ref{fig:vr_lambda_d} but for the zenithal angle of the Evershed flow. The heights are from model D.}
\label{fig:theta_lambda_d}
\end{figure*}

Finally, Figure \ref{fig:theta_lambda_d}  shows the inclination angle of the Evershed flow given by the ratio of the vertical and horizontal velocities. The calculation of the angle is largely affected by the uncertainties in the determination of the vertical velocities. No dependences can be detected. In all the cases, the difference between the angle of the neutral and ionized flow is within the error bar limits. 

\section{Discussion of the results and uncertainties}

Above we have presented evidences that the radial component of the velocity of the Evershed flow measured from neutral iron lines is larger than that measured from singly ionized iron lines. The differences are systematic and are present at all heights and radial distances of the penumbra, for all five pairs of the observed spectral lines. The uncertainties caused by the larger errors of the vertical velocity determination due to the limb location of the observed spot do not allow us to conclusively measure this component, neither the zenithal angle of the Evershed flow. Nevertheless, the radial component of the flow is determined rather reliably and the difference between the neutral and ion velocities of a few hundred \ms\ is above the error bar limits. 

In our analysis we excluded the uncertainties due to the absolute wavelength calibration, since in all cases pairs of spectral lines are observed with the same CCD. We have also corrected for the velocity gradient coming from the curvature of spectral lines caused by the spectrograph, and for the gradient caused by solar rotation in the field of view at the time of the scan. An additional correction of the $\lambda$-meter velocities for the bisector shape of the spectral lines was also applied.  Altogether these corrections largely remove possible systematic errors in the velocity measurements. 

The observed difference in velocities is not large. As far as we aware of, this is the first time such a difference is detected. Due to the magnitude of the difference, this comes with no surprise.  Measurements of the Evershed flow from neutral and iron lines done independently would not be able to reveal it, since the variations of the Evershed flow from one sunspot to another, and even over the same sunspot at different times, may be larger than the difference between the ion and neutral velocities we have found. Our analysis is also possible due to a unique dataset taken under extremely good and stable seeing conditions.

The main source of uncertainty in our analysis comes from the determination of heights where the velocity signal comes from. In our recent work \citep{Diaz+etal2015}, we have measured ion and neutral velocities in a prominence located at the solar limb. In that case, since the prominence material is essentially optically thin, the signal can reasonably be assumed to originate from the same region of the prominence plasma. Unlike that, the formation height of iron spectral lines in the penumbra spans a range of photospheric heights. These heights can vary depending on the assumed thermodynamic structure of the penumbral atmosphere. Therefore, assigning formation heights is crucial in our analysis. 

As mentioned in the introduction, the Evershed flow is observed to be directed along the more inclined intra-spines, that coincide with dark filaments in the outer penumbra and bright filaments in the inner penumbra \citep{Schmidt+Schlichenmaier2000, Schlichenmaier+etal2005, BellotRubio+etal2006, Ichimoto+etal2007}.  Since dark material usually mean cooler and therefore more neutral plasma, one may conjecture that neutral iron lines mainly form in dark filaments (those harboring Evershed flow in the outer penumbra) and ionized iron lines form in the hotter and brighter filaments with stronger and more vertical field. This may partially or fully explain larger amplitudes of the Evershed flow detected from neutral lines in our observations. In order to check if this effect is operating, we made an experiment using heights from model D (bright filament) for ionized lines and heights from model C (dark filament) for neutral lines in presenting the dependences for the radial component of the Evershed flow (as in Figures \ref{fig:vr_lambda_c} and \ref{fig:vr_lambda_d}). We found that qualitatively, all the dependences remain the same. 

To perform an additional check, we have synthesized the observed spectral lines using both models C and D, introducing a velocity field that bright and dark filaments are expected to harbor. We assumed a velocity linearly decreasing with height by about 0.3 \kms\ in 100 km starting from a fixed photospheric value, the latter one taken to be different in the model C and D. We made several experiments in which we varied the photospheric velocity values in the model C between 0 and 2 \kms\ and the one in the model D from 3 to 6 \kms.
We then produced an average profile of each line using a combination of the profiles from both models with a different weight and repeated the $\lambda$-meter procedure to extract velocities from the mixed synthetic profiles. Our experiment revealed that, independently of the weight of the models, the $\lambda$-meter velocities obtained from neutral and ionized lines are very similar, with differences much smaller than those found in our observations.  Only in the case the difference in velocities between model C and D exceeds 3 \kms\ the $\lambda$-meter velocities measured from synthetic neutral lines become slightly higher than those from ionized lines, for some pairs of lines. The difference does not exceed 100 \ms\ on average and is smaller than the observed one.  \citet{BellotRubio+etal2004} give the Evershed flow velocities extracted by inversion of two IR \FeI\ lines at 1.56 $\mu$m using a two-component model of penumbra (see their Figure 4), corresponding to the deep photosphere. According to their results, in the outer penumbra the velocity from the more vertical component reaches about 2 \kms, while the one from the more horizontal component reaches 3-4 \kms. Therefore, we conclude that only a small part of the observed difference can be accounted for by the above effect.

Yet another argument comes from the fact that the difference between \FeI\ and \FeII\ velocities is qualitatively similar for all distances in the penumbra. In the outer penumbra the Evershed flow is assumed to be harbored by dark and cool filaments, but in the inner penumbra it is along bright and hotter filaments. Nevertheless, we find no reversal in the behavior of \FeI\ and \FeII\  velocities in Figures \ref{fig:vr_lambda_c} and \ref{fig:vr_lambda_d}.

\citet{Ichimoto1988} provided a thorough study of the Evershed flow dependence on various parameters of spectral lines for a set containing 85 simultaneously observed lines of different elements with different ionization potentials, sensitivities to the magnetic field, and ionization stages.  He found that the Evershed flow speed measured in the wings of spectral lines increases with increasing excitation potential. We find no such dependence in our case, given that the range of excitation potential of the sample used in the present study is much narrower (2-4 eV compared of 0.5 - 6 eV in \citet{Ichimoto1988}). It is also briefly mentioned by \citet{Ichimoto1988} that no significant difference between the velocity of the neutral and ionized lines was found. However, Figure 2 of this paper shows a slight tendency of ionized species to have a lower velocity than neutral atoms, indicating the same behavior as found in this work. \citet{Ichimoto1988} performed simulations of the artificial \FeI\ and \FeII\ lines profiles using different hypothesis about the temperature, density and velocity stratifications in the bright and dark penumbral filaments. He found that, if dark filaments are assumed to be cooler, velocities from \FeI\ lines are larger than those from \FeII\ lines, but the observed dependence with the excitation potential can not be reproduced. Such dependence is only reproduced when the dark filaments are assumed to be denser, but with a similar temperature as bright filaments. This gives us another evidence that differences in the \FeI\ and \FeII\ line velocities found in our work are not caused by the temperature-velocity correlations in the penumbral filaments.

Alternatively, one can speculate that the observed difference in the neutral and ion velocities can be caused by an incomplete coupling between the different plasma components in the strongly magnetized sunspot atmosphere. In the photosphere the collisional coupling of plasma is very strong. Nevertheless, as is mentioned in the Introduction, the dominance of the neutral atoms at photospheric heights can lead to a partial decoupling between the neutral and the ionized components. The degree of these decoupling depends on the magnetic field and can be expected to be stronger in the sunspot atmosphere with a larger field and cooler plasma. In  order to check if this effect could be at work, we used the 3D models of sunspot obtained from inversion of IR Ca triplet by \citet{Socas-Navarro2005} and calculated the drift velocity $\vec{w}=\vec{u}_i -\vec{u}_n$ (relative ion-neutral velocity) from the parameters of the model using the equation from \citet{Khomenko+etal2014b}:

\begin{equation}
\label{eq:w_bis}
\vec{w} = \frac{\xi_n}{\alpha_n} \left[\vec{J} \times\vec{B} \right] - \frac{\vec{G}}{\alpha_n} + m_e\nu_{en}\frac{\vec{J}}{e \alpha_n}
\end{equation}
where $\xi_n$ is the neutral fraction; $\alpha_n$ is the sum of collisional frequencies between neutrals and other species multiplied by the corresponding mass densities; $\nu_{en}$ is the electron-neutral collisional frequency; $\vec{B}$ is the magnetic field vector, $\vec{J}$ is the current; $m_e$ and $e$ are the electron mass and charge. The quantity $\vec{G}=\xi_n \vec{\nabla}{\bf{p}_{e}} - (1-\xi_n) \vec{\nabla}  {\bf{p}_n}$ is the partial pressure gradient term, being $p_e$ the electron pressure and $p_n$ the neutral pressure. Note that there are several effects that contribute into \vec{w}, and it is difficult to guess from general considerations the sign of this velocity. All the quantities entering Eq. \ref{eq:w_bis} were extracted or calculated from the parameters of the model by  \citet{Socas-Navarro2005}. This calculation has to be taken with care because of the relatively low spatial resolution in the derivatives (200 km) and all other approximations assumed when doing the inversion of the IR Ca triplet profiles \citep[see][]{Socas-Navarro2005}.  Special care must be taken when doing horizontal derivatives since different pixels in the semi-empirical model are set to a common geometrical height only approximately. Nevertheless it might give an idea of the order of magnitude of the drift velocity to be expected in the sunspot penumbra. The horizontal component of the drift velocity results to be larger than the vertical component and therefore should be better detected in observations done near the limb, as in our case. Curiously, the term that contributes most to this velocity is the partial pressure gradient term $\vec{G}$, and the terms related to currents are orders of magnitude smaller. This can be due to the relatively low resolution of the model. Given all the uncertainties, this calculation is not able to provide us a clue of the sign of the difference.

Our calculation shows that the magnitude of the velocity increases with height, in agreement with the decrease of the collisional coupling (the value of $\alpha_n$). Surprisingly, even in the photosphere at some locations it reaches the measurable values of 10$^2$ \ms, while in the chromosphere it can increase up to \kms\ values. About 0.4\% of the field of view in the photosphere, and 6 \% in the chromosphere has drift velocities in the range of 1-100 \ms. Therefore, is can not be excluded that measurable drift velocity can be present in sunspots and that part of the observed ion-neutral velocity difference detected in this work can be attributed to the decoupling of the different plasma components. Ideally, chromospheric lines and specially designed observing campaigns should be better used in order to detect such decoupling in future studies.

\begin{acknowledgements}
This work is partially supported by the Spanish Ministry of Science through projects AYA2010-18029, AYA2011-24808 and AYA2014-55078-P. This work contributes to the deliverables identified in FP7 European Research Council grant agreement 277829, ``Magnetic connectivity through the Solar Partially Ionized Atmosphere''. NS is grateful to the Spanish Ministry of Science which granted her sabbatical stay at the IAC through the 2011 Severo Ochoa Program SEV-2011-0187.
\end{acknowledgements}
%
\providecommand{\noopsort}[1]{}\providecommand{\singleletter}[1]{#1}

\end{document}